\documentclass[aps,prb,twocolumn,superscriptaddress,showpacs]{revtex4-1}

\usepackage{graphicx}
\usepackage{calc}
\usepackage{bm}
\usepackage{color}
\usepackage{textcomp}

\begin{document}

\title{Second-harmonic voltage responce for the magnetic Weyl semimetal Co$_3$Sn$_2$S$_2$}

\author{V.D.~Esin}
\author{A.V.~Timonina}
\author{N.N.~Kolesnikov}
\author{E.V.~Deviatov}
\affiliation{Institute of Solid State Physics of the Russian Academy of Sciences, Chernogolovka, Moscow District, 2 Academician Ossipyan str., 142432 Russia}

\date{\today}

\begin{abstract}
    We experimentally investigate longitudinal and transverse second-harmonic voltage response to ac electrical current for a magnetic Weyl semimetal  Co$_3$Sn$_2$S$_2$. In contrast to the previously observed Berry-curvature induced non-linear Hall effect for non-magnetic Weyl and Dirac semimetals, the second-harmonic transverse voltage demonstrates sophisticated interplay of different effects for Co$_3$Sn$_2$S$_2$. In high magnetic fields, it is of Seebeck-like  square-B law, while the low-field behavior is found to be linear and sensitive to the direction of sample magnetization. The latter can be expected both  for the non-linear Hall effect and for the surface state contribution to the Seebeck effect in Weyl semimetals. Thus, thermoelectric effects are significant in  Co$_3$Sn$_2$S$_2$, unlike non-magnetic Weyl and Dirac materials.
\end{abstract}

\pacs{73.40.Qv  71.30.+h}

\maketitle


Recent interest to the time-reversal-invariant non-linear Hall (NLH) effect~\cite{sodemann,nlhe1,nlhe2,nlhe3,nlhe4,nlhe5,nlhe6,nlhe7,nlhe8,nlhe9,nlhe10,nlhe11,nlhe12,nlhe13}  is a part of a broad research area of topological systems. In zero magnetic field, a non-linear Hall-like current  arises from the Berry curvature, which can be be regarded as a magnetic field  in momentum space. It leads to a 
quadratic response to ac excitation current, so NLH effect should appear as a non-zero transverse second-harmonic voltage without magnetic field. Since Berry curvature concentrates in regions  where two or more bands cross~\cite{armitage}, topological systems are the obvious candidates to observe the NLH effect~\cite{sodemann}.  It has been experimentally demonstrated for monolayer transitional metal dichalcogenides~\cite{ma,kang} and for three-dimensional Weyl and Dirac semimetals~\cite{esin}. 

Dirac semimetals  host special points of Brillouin zone with three dimensional linear dispersion~\cite{armitage}. In Weyl semimetals, by breaking time reversal or inversion symmetries, every Dirac point splits  into two Weyl nodes with opposite chiralities.  First experimentally investigated WSMs  were non-centrosymmetric crystals with broken inversion symmetry.  Even in this case, a second-harmonic  quadratic signal can also originate from the thermoelectric Seebeck effect~\cite{ptsn,cdas_thermo}.   When the magnetic field is perpendicular to the temperature gradient,  it leads to  quadratic-B correction in the Seebeck coefficient~\cite{thermo_theor1,thermo_theor2}. In contrast to these calculations, a second-harmonic NLH voltage shows~\cite{esin} odd-type dependence on the direction of the magnetic field. Thus, the magnetic field measurements allow to distinguish the NLH effect from the thermoelectric response.

There are only a few candidates~\cite{mag1,mag2,mag3,mag4} of magnetically ordered  WSMs with broken time-reversal symmetry. Recently,  giant anomalous Hall effect was reported~\cite{kagome,kagome1} for the kagome-lattice ferromagnet Co$_3$Sn$_2$S$_2$, as an indication for the existence of a magnetic Weyl phase. Sophisticated regimes of second-harmonic response should also be expected  in  Weyl semimetals with broken time reversal symmetry. In addition to the expected Berry curvature contribution to the Hall-like currents, the chiral anomaly contribution to second-harmonic generation in the lowest order is linearly proportional to the applied magnetic field~\cite{zyuzin}. Moreover, in magnetic materials, Nernst voltage can be generated normally to the temperature gradient even without an external magnetic field, which is known as anomalous Nernst effect (ANE). ANE was reported for different Co$_3$Sn$_2$S$_2$ thermoelectric  devices~\cite{kagome_Nernst1,kagome_Nernst2,kagome_Nernst3}.

Here, we experimentally investigate longitudinal and transverse second-harmonic voltage response to ac electrical current for a magnetic Weyl semimetal  Co$_3$Sn$_2$S$_2$. In contrast to the previously observed Berry-curvature induced non-linear Hall effect for non-magnetic Weyl and Dirac semimetals, the second-harmonic transverse voltage demonstrates sophisticated interplay of different effects for Co$_3$Sn$_2$S$_2$. In high magnetic fields, it is of Seebeck-like  square-B law, while the low-field behavior is found to be linear and sensitive to the direction of sample magnetization. The latter can be expected both  for the non-linear Hall effect and for the surface state contribution to the Seebeck effect in Weyl semimetals. Thus, thermoelectric effects are significant in  Co$_3$Sn$_2$S$_2$, unlike non-magnetic Weyl and Dirac materials.


Co$_3$Sn$_2$S$_2$ single crystals were grown by the gradient freezing method, see Refs.~\onlinecite{cosns,cosnsjc} for details. The kagome-lattice ferromagnet Co$_3$Sn$_2$S$_2$ can be easily cleaved, the Laue patterns confirm the hexagonal structure with $(0001)$ as cleavage plane. Magnetoresistance measurements confirms high quality of our  crystals:  Co$_3$Sn$_2$S$_2$ samples  demonstrate~\cite{cosns,cosnsjc} giant anomalous Hall effect and positive, non-saturating longitudinal magnetoresistance, which even quantitatively coincide with the previously reported ones~\cite{kagome,kagome1}.  

\begin{figure}
\centerline{\includegraphics[width=0.8\columnwidth]{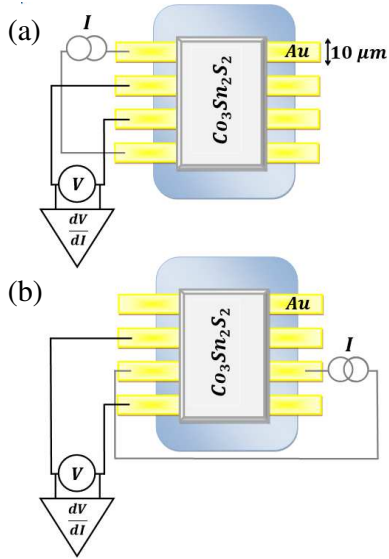}}
\caption{(Color online) Sketch of the sample with electrical connections.  Au leads are formed on a SiO$_2$ substrate, with 5~$\mu$m intervals between them. A thick Co$_3$Sn$_2$S$_2$ flake ($\approx$ 100~$\mu$m size)  is transferred~\cite{cosns,cosnsjc} on the top of the leads, forming small Ohmic contacts ($\approx$ 10 $\mu$m overlap between the Co$_3$Sn$_2$S$_2$ flake and the leads). Current flows  along the sample edge in (a) for $V^{2\omega}_{xx}(I)$ investigations, and normally to it in (b) for $V^{2\omega}_{xy}(I)$ ones. The second-harmonic (2$\omega$) component of the longitudinal voltage $V^{2\omega}_{xx}(I)$  is measured in a standard four-point lock-in technique.}
\label{sample}
\end{figure}

Despite it is possible to form contacts directly on the  cleaved Co$_3$Sn$_2$S$_2$ crystal plane~\cite{cdas}, it is not desirable for transport investigations: the leads to the bonding pads would also participate in current distribution. To obtain a definite sample geometry, the leads pattern is formed on the insulating SiO$_2$ substrate by lift-off technique after thermal evaporation of 100~nm Au, see Fig.~\ref{sample}. Small (about 100~$\mu$m size and 1~$\mu$m thick) Co$_3$Sn$_2$S$_2$ flakes are obtained by a mechanical cleaving method. The most plane-parallel flakes  with clean surface are transferred to the Au leads pattern, see Refs.~\cite{cosns,cosnsjc} for details.  This procedure provides reliable Ohmic contacts, stable in multiple cooling cycles, which has been successfully demonstrated for different layered materials~\cite{cosns,cosnsjc,cdas,inwte}.

We measure the second-harmonic longitudinal $V^{2\omega}_{xx}(I)$ and transverse $V^{2\omega}_{xy}(I)$ voltage components in standard four-point lock-in technique, see the principal circuit diagrams in Fig.~\ref{sample} (a) and (b), respectively.  The potential contacts are always situated along the sample edge, while the ac current $I$ flows  along the edge for $V^{2\omega}_{xx}(I)$ investigations in (a), and normally to it in (b) for $V^{2\omega}_{xy}(I)$ ones. 

We ensure, that the measured second-harmonic $V^{2\omega}$ voltage is antisymmetric with respect to the voltage probe swap and it is also independent of the ground probe position. We check, that the lock-in signal is not sensitive to the ac current frequency in the range 1100 Hz -- 1kHz, which is defined by applied filters. 
For Co$_3$Sn$_2$S$_2$, magnetic moments order ferromagnetically~\cite{kagome} below 175~K. For this reason, the measurements are performed in a usual He4 cryostat equipped with a superconducting solenoid. We do not obtain noticeable temperature dependence in the interval 1.4-4.2~K, so all the results below are obtained at 4.2~K and for the normal (to the flake's plane) magnetic field orientation.


\begin{figure}
\includegraphics[width=\columnwidth]{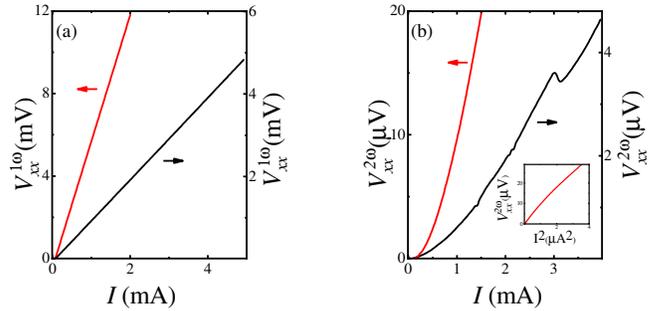}
\caption{(Color online) Examples of the longitudinal first- $V^{1\omega}_{xx}(I)$ (a) and second-harmonic $V^{2\omega}_{xx}(I)$ (b) characteristics for two different samples of a magnetic Weyl semimetal Co$_3$Sn$_2$S$_2$. For strictly linear first-harmonic $V^{1\omega}_{xx}(I)$ in (a), one can not expect~\cite{ma,kang,esin} non-zero  second-harmonic $V^{2\omega}_{xx}$ signal in (b).  The significant $V^{2\omega}_{xx}(I)$ should be connected with the anomalous Nernst effect~\cite{kagome_Nernst1,kagome_Nernst2,kagome_Nernst3}, since it  is normal both to the temperature gradient and to the Co$_3$Sn$_2$S$_2$ magnetization direction in Fig.~\ref{sample}. Inset confirms the quadratic dependence $V^{2\omega}_{xx}(I) \sim I^2$, which well corresponds to the Joule heating effect.  The curves are obtained at 4.2~K temperature in zero magnetic field.}
\label{IV}
\end{figure}

Examples of the first-harmonic longitudinal voltage $V^{1\omega}_{xx}$ are shown as function of ac current $I$ in Fig.~\ref{IV} (a) for two different samples in zero magnetic field. The curves are strictly linear, they correspond to 6~Ohm and 2~Ohm sample's resistance.  In this case,  second-harmonic signal $V^{2\omega}_{xx}(I)$ can not be produced by $I-V$ non-linearity.  In contrast to this expectation, we observe significant second-harmonic longitudinal voltage for these two samples, see Fig.~\ref{IV} (b). $V^{2\omega}_{xx}(I)$ is strongly non-linear, it is proportional to the square of the applied current, as it is demonstrated in the inset to Fig.~\ref{IV} (b). The obtained $V^{2\omega}_{xx}(I)$ values are in four times higher for the resistive (6~Ohm) sample in comparison with the other (2~Ohm) one.

This behavior strongly contradicts to zero $V^{2\omega}_{xx}(I)$ for non-magnetic monolayer transitional metal dichalcogenides~\cite{ma,kang} and for three-dimensional Weyl and Dirac semimetals~\cite{esin}.  The inherent magnetization of a thick Co$_3$Sn$_2$S$_2$ flake is perpendicular to the flake’s plane~\cite{kagome}, so non-zero $V^{2\omega}_{xx}(I)$ should be connected with the anomalous Nernst effect~\cite{kagome_Nernst1,kagome_Nernst2,kagome_Nernst3} for the magnetic Weyl semimetal Co$_3$Sn$_2$S$_2$ in Fig.~\ref{IV} (b).  Temperature gradient is created in the Co$_3$Sn$_2$S$_2$ flake  due to the Joule heating by ac current $I$, the gradient is perpendicular to the current line in Fig.~\ref{sample} (a).  Nernst voltage appears as longitudinal $V^{2\omega}_{xx}(I)$ in our experimental setup, since it  is normal both to the temperature gradient and to the magnetization direction. The experimental observation  $V^{2\omega}_{xx}(I)\sim I^2$ also well corresponds to the Joule heating effect.

\begin{figure}
\centerline{\includegraphics[width=\columnwidth]{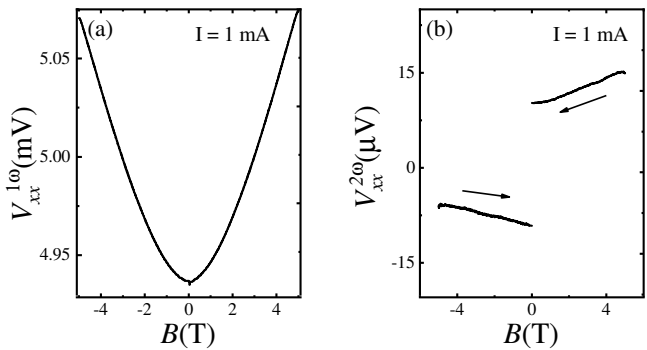}}
\caption{(Color online) Magnetic field dependence of first- (a)  and second-harmonic (b) longitudinal voltage $V_{xx}$  at fixed ac current $I=1$~mA. To avoid magnetization switchings~\cite{kagome,kagome1,cosns} in low fields, the  curves are obtained by sweep from highest field values for both fields' directions. 
In (a), $V^{1\omega}_{xx}(B)$ is an even function of $B$, it demonstrates usual non-saturating longitudinal magnetoresistance~\cite{kagome,kagome1,cosns} in Co$_3$Sn$_2$S$_2$.   In (b), two $V^{2\omega}_{xx}(B)$ branches reflect anomalous Nernst voltage of different sign due to the opposite magnetization directions for these branches.  All the curves are obtained at 4.2~K temperature. 
}
\label{field}
\end{figure}

It has been demonstrated~\cite{esin} for non-magnetic three-dimensional Weyl and Dirac semimetals, that magnetic field measurements are important to establish an origin of the second-harmonic voltage.  First- and second-harmonic $V_{xx}(B)$ dependencies are shown in Fig.~\ref{field}, (a) and (b), respectively. The first-harmonic component $V^{1\omega}_{xx}(B)$ is an even function of $B$, it demonstrates usual non-saturating longitudinal magnetoresistance~\cite{kagome,kagome1,cosns} in normal magnetic field. The $V^{2\omega}_{xx}(B)$ dependence is a nearly odd function, there is also a significant jump in zero magnetic field, like it is expected for ANE~\cite{kagome_Nernst1,kagome_Nernst2,kagome_Nernst3}. 

The zero-field jump can be understood as a result of sample magnetization switching.  To work with a definite magnetic state of Co$_3$Sn$_2$S$_2$, the  curves are obtained in Fig.~\ref{field} by  sweep from highest field values both for the positive and for the negative fields. Because Co$_3$Sn$_2$S$_2$   demonstrates giant anomalous Hall effect~\cite{kagome,kagome1,cosns}, the magnetization directions are opposite in zero field for these two curves in Fig.~\ref{field}, (b). Thus, the curves  correspond to ANE signals of different sign, which we observe as the zero-field $V^{2\omega}_{xx}(I)$ jump.  On the other hand, any imaginary $I-V$ nonlinearity effects should be symmetric in magnetic field, due to the even $V^{1\omega}_{xx}(B)$ magnetoresistance dependence in Fig.~\ref{field} (a), so the magnetic field behavior confirms thermoelectric origin of finite longitudinal second-harmonic $V^{2\omega}_{xx}(I)$ voltage.

\begin{figure}
\includegraphics[width=\columnwidth]{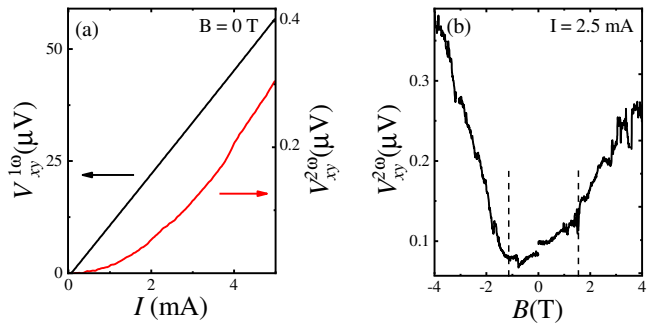}
\caption{(Color online) (a) First- $V^{1\omega}_{xy}(I)$ and second-harmonic $V^{2\omega}_{xy}(I)$ transverse voltage components in zero magnetic field. The linear Hall component $V^{1\omega}_{xy}(I)\sim I$ reflects finite~\cite{cosns} Co$_3$Sn$_2$S$_2$ magnetization. The nonlinear second-harmonic $V^{2\omega}_{xy}(I)$  is one magnitude smaller than  for the $xx$ configuration in Fig.~\ref{IV}. (b)  $V^{2\omega}_{xy}(B)$ is always positive for both fields' directions, it demonstrates strong nonlinear increase in high magnetic fields. However,  $V^{2\omega}_{xy}(B)$ is obviously not symmetric, there is also a linear region between -1.1~T and +1.7~T with a small zero-field jump.  Both $V^{2\omega}_{xy}(B)$  branches are obtained by  sweep from the highest fields. All the curves are presented for 4.2~K temperature.
}
\label{IVxy}
\end{figure}

Similar behavior can be demonstrated for the transverse $V_{xy}(I)$ voltage component in zero magnetic field,  see Fig~\ref{IVxy} (a). The linear Hall voltage $V^{1\omega}_{xy}(I)\sim I$ is due to the finite Co$_3$Sn$_2$S$_2$ magnetization in zero external field. We also obtain non-linear second-harmonic $V^{2\omega}_{xy}(I)\sim I^2$, which is one magnitude smaller than  for the $xx$ configuration in Fig.~\ref{IV} (b), despite of similar first-harmonic values in Figs.~\ref{IV} (a) and~\ref{IVxy} (a . In principle, finite $V^{2\omega}_{xy}(I)\sim I^2$ can be produced~\cite{esin} both by NLH and by the thermoelectric effects. In the latter case, the second-harmonic voltage reflects the Seebeck effect~\cite{esin}, because  potential contacts are parallel to the temperature gradient in Fig.~\ref{sample} (b). 

For non-magnetic Weyl and Dirac semimetals, $V^{2\omega}_{xy}(B)$ behaves as an odd function of magnetic field due to NLH effect, which has been demonstrated for the exactly the same  experimental geometry~\cite{esin}.  In contrast, we observe sophisticated magnetic field behavior in Fig~\ref{IVxy} (b), despite the samples are prepared in the same technique and the contact configuration is also the same~\cite{esin}.  $V^{2\omega}_{xy}(B)$ is always positive for both fields' directions, it demonstrates strong nonlinear increase in high magnetic fields. However,  $V^{2\omega}_{xy}(B)$ is obviously not symmetric, there is also a linear region between -1.1~T and +1.7~T with a small zero-field jump. 

As a result, the longitudinal second-harmonic voltage $V^{2\omega}_{xx}$ behavior well corresponds to the expected one for the anomalous Nernst effect in our experimental setup. In contrast, the magnetic field behavior $V^{2\omega}_{xy}(B)$ is quite sophisticated, it strongly contradicts to the known one  both for NLH and for the Seebeck effects in non-magnetic materials~\cite{esin}. 

Let us start from the thermoelectric effects in Weyl and Dirac semimetals. When the magnetic field is perpendicular to the temperature gradient, the longitudinal thermal conductivity (Seebeck effect) is expected~\cite{thermo_theor1,thermo_theor2} to pick up a negative contribution that goes as the square of the magnetic field. In our experimental setup, the longitudinal thermal conductivity  corresponds to the inverse $V^{2\omega}_{xy}(B)$ value, as described above,  so the high-field behavior in Fig~\ref{IVxy} (b) is in agreement with the theoretical predictions~\cite{thermo_theor1,thermo_theor2}. 

On the other hand, the low-field linear behavior with the zero-field jump demands another explanation. On the one hand, Berry curvature dipole NLH effect~\cite{sodemann,nlhe1,nlhe2,nlhe3,nlhe4,nlhe5,nlhe6,nlhe7,nlhe8,nlhe9,nlhe10,nlhe11,nlhe12,nlhe13} should also be seen in magnetically ordered WSMs. Berry curvature acts analogously to a magnetic field in the momentum space, so NLH voltage is linear in external magnetic field~\cite{esin} and also picks up the Co$_3$Sn$_2$S$_2$ magnetization. The latter should lead to the zero-field jump, since the $V^{2\omega}_{xy}(B)$  branches correspond to different magnetization directions in Fig~\ref{IVxy} (b). On the other hand, one can expect some contribution from Fermi arcs to magnetothermal transport in Weyl semimetals. For topological insulators the latter effect is known to produce large and anomalous Seebeck effects with an opposite sign to the Hall effect~\cite{thermo_ES}. Fermi arcs were directly demonstrated for Co$_3$Sn$_2$S$_2$  by scanning tunneling spectroscopy~\cite{kagome_arcs}.

These two possibilities could be distinguished by dependence of $V^{2\omega}_{xy}(B)$ on the magnetic field orientation~\cite{thermo_theor1,thermo_theor2}. However, Co$_3$Sn$_2$S$_2$ magnetic properties arise from the kagome-lattice cobalt planes, whose magnetic moments order ferromagnetically~\cite{kagome} out of plane below 175~K. We have tested, that even comparatively small parallel magnetic field component results in the force, which is large enough to detach the Co$_3$Sn$_2$S$_2$ flake from the substrate.

As a conclusion, we experimentally investigate longitudinal and transverse second-harmonic voltage response to ac electrical current for a magnetic Weyl semimetal  Co$_3$Sn$_2$S$_2$. We find, that the longitudinal component depends quadratically on the ac current in zero magnetic field, which reflects strong anomalous Nernst effect, well known  for  Co$_3$Sn$_2$S$_2$ material. In contrast to the previously observed Berry-curvature induced non-linear Hall effect for non-magnetic Weyl and Dirac semimetals, the second-harmonic transverse voltage demonstrates sophisticated interplay of different effects for Co$_3$Sn$_2$S$_2$. In high magnetic fields, it is of Seebeck-like  square-B law, while the low-field behavior is found to be linear and sensitive to the direction of sample magnetization. The latter can be expected both  for the non-linear Hall effect and for the surface state contribution to the Seebeck effect in Weyl semimetals. Thus, thermoelectric effects are significant in  Co$_3$Sn$_2$S$_2$, unlike non-magnetic Weyl and Dirac materials.

\acknowledgments
We wish to thank  V.T.~Dolgopolov for fruitful discussions, O.O. Shvetsov for samples preparation.  We gratefully acknowledge financial support partially by the RFBR  (project No.~19-02-00203), RAS, and RF State task.

\end{document}